\documentclass[12pt]{amsart}
\usepackage{amssymb}
\usepackage{verbatim}
\usepackage[usenames]{color}
\usepackage{hyperref}

\newtheorem{thm}{Theorem}
 
\newtheorem{la}[thm]{Lemma}

\theoremstyle{definition}

\theoremstyle{remark}

\renewcommand{\phi}{\varphi}

\newcommand{\notarrow}{\kern .42em\not\kern -.42em\longrightarrow}

\newcommand{\noprint}[1]{\relax}

\title[Rates of Minimal Non-Matroid Access Structures]{Information Rates of Minimal Non-Matroid-Related Access Structures}
\author{Jessica Ruth Metcalf-Burton}
\address{Mathematics Department\\
University of Michigan\\
Ann Arbor, MI 48109--1043, U.S.A.}
 \email{jmetcalf@umich.edu}

\begin{document}

\begin{abstract}

In a secret sharing scheme, shares of a secret are distributed to participants in such a way that only certain predetermined sets of participants are qualified to reconstruct the secret.  An access structure on a set of participants specifies which sets are to be qualified.  The information rate of an access structure is a bound on how efficient a secret sharing scheme for that access structure can be.
Mart\'i-Farr\'e and Padr\'o showed that all access structures with information rate greater than $\frac{2}{3}$ are matroid-related, and Stinson showed that four of the minor-minimal, non-matroid-related access structures have information rate exactly $\frac{2}{3}$.  By a result of Seymour, there are infinitely many remaining minor-minimal, non-matroid-related access structures.  These are of the form $\Gamma_n$ for $n \geq 4$ , where
\begin{equation*}\Gamma_n = \{ \{k,p_i\} | 1 \leq i \leq n \} \cup \{\{p_1, \dots , p_n\} \}\end{equation*}
is the access structure on a king and $n$ pawns.
 In this paper we find the exact information rates for all $\Gamma_n$, $n \geq 2$, thus making known the information rates for all minimal non-matroid-related access structures. 

\end{abstract}

\maketitle

\section{Background and Motivation}

Our treatment of secret sharing follows that of Csirmaz \cite{Csirmaz} and Mart\'{i}-Farr\'{e} and Padr\'{o} \cite{MF}.
Let $P$ be a set of participants, among whom we would like to share a secret.     An {\it access structure} $\Gamma$ on $P$ is the collection of all 
subsets of $P$ that are {\it qualified}, i.e., allowed to reconstruct the secret.  An access structure 
$\Gamma$ is fully determined by its {\it minimal qualified subsets}, which are those qualified sets for which no proper subset is qualified.  Any subset of $P$ not in $\Gamma$ 
is called {\it unqualified}.  We assume that each participant in $P$ belongs to some minimal qualified subset.

Intuitively, a secret sharing scheme for $\Gamma$ is a way of taking a secret and using it to assign
one or more {\it shares} to each participant
in such a way that qualified sets are able to reconstruct the secret by combining their shares, while unqualified sets cannot learn any information about the value of the secret.  

More precisely, let $\Sigma$ be a collection of random variables consisting of one for the secret and one for each
participant in $P$.  We use $S$ to denote both the secret and its associated random variable.
For any participant $x \in P$, we use $H(x)$ to denote the Shannon entropy of the corresponding 
random variable, and for any
subset $X \subseteq P \cup \{S\}$, we use $H(X)$ to denote the joint entropy of the 
random variables for all elements of $X$.  We use $H(X|Y)$ similarly for conditional entropy when 
$X,Y \subseteq P \cup \{S\}$.

We call $\Sigma$ a {\it (perfect) secret sharing scheme for $\Gamma$} if it has the following properties:
\begin{itemize}
\item If $X \in \Gamma$ then  $H(S|X)=0$, that is, the participants in $X$ are able to combine their shares to 
completely determine the value of the secret.
\item If $X \notin \Gamma$ then $H(S|X)=H(S)$, that is, the uncertainty about the secret does not change even when all participants in $X$ pool their shares.
\end{itemize}

One measure of the efficiency of a secret sharing scheme is its 
{\it information rate}.
Given a secret sharing scheme $\Sigma$ and a participant $x \in P$, the information rate of $x$ is defined by
\begin{equation*} \rho(x)=\frac{H(S)}{H(x)} .\end{equation*}
The information rate of $\Sigma$, $\rho(\Sigma)$, is the minimum information rate over all participants in $P$.
For an access structure $\Gamma$, the information rate $\rho(\Gamma)$ is the supremum of $\rho(\Sigma)$ over all
$\Sigma$ that are secret sharing schemes for $\Gamma$.

The information rate of a participant, secret sharing scheme, or access structure must always be less than or equal to one \cite{Csirmaz}.  It has been shown that all access structures with information rate greater than $\frac{2}{3}$ are {\it matroid-related}.  For this result, as well as a definition and further details about matroid-related access structures,
we refer the reader to Mart\'i-Farr\'e and Padr\'o \cite{MF}.  

An interesting class of access structures that are not matroid-related are the {\it king and $n$ pawns} access structures $\Gamma_n$ for $n \geq 3$.
$\Gamma_n$ is defined on the set of participants $P_n=\{k, p_1, \dots, p_n\}$ by
\begin{equation*}\Gamma_n = \{ \{k,p_i\} | 1 \leq i \leq n \} \cup \{\{p_1, \dots , p_n\} \}.\end{equation*}
In $\Gamma_n$ the king and any one pawn may reconstruct the secret, as may the set of all pawns.  However, the king may not reconstruct the secret alone, nor may any proper subset of the pawns.  

Seymour  \cite{Seymour} showed that the minor-minimal access structures that are non-matroid-related are $\{\Gamma_n\}_{n\geq 3}$ along with $\{\{a,b\},\{b,c\},\{c,d\}\}$, $\{ \{a,b\}, \{a,c\}, \{a,d\}, \{b,c\} \}$, and $\{ \{a,b\}, \{a,c\}, \{b,c,d\}\}$.  Stinson \cite{Stinson} demonstrated that these last three access structures, as well as $\Gamma_3$, all have information rates of $\frac{2}{3}$.   
To the best of the author's knowledge, the information rates of the structures $\Gamma_n$ have not been determined for general $n$.   In this paper we remedy this, by finding the exact information rates for all structures in the infinite class $\{\Gamma_n\}_{n \geq 2}$.

\section{Results}

Our main result  is 

\begin{thm}
For $n \geq 2$, \begin{equation*}\rho(\Gamma_n)=\frac{n-1}{2n-3}.\end{equation*}
\end{thm}

We will prove this theorem by showing that $\frac{n-1}{2n-3}$ is both an upper and a lower bound for the information rate of $\Gamma_n$.
 
While proving bounds on the information rate, we will use the normalized entropy function $h$, defined by
\begin{eqnarray*}h(X)=\frac{H(X)}{H(S)}\end{eqnarray*}
for any subset $X \subseteq P \cup \{S\}$.
Observe that for a singleton $\{x\}$, $h(\{x\})$ is the reciprocal of $\rho(x)$.
The function $h$ has the following properties, proofs of which may be found in Csirmaz \cite{Csirmaz}:
\begin{itemize}
\item $h(\emptyset)=0$.

\item Monotonicity: For $X \subseteq Y \subseteq P \cup \{S\}$, 
\begin{equation*}h(X) \leq h(Y).\end{equation*}

\item Submodularity: For $X,Y \subseteq P \cup \{S\}$, 
\begin{equation*}h(X) + h(Y) \geq h(X \cap Y) + h(X \cup Y).\end{equation*}

\item $+$-Submodularity: For $X,Y \in \Gamma$ but $X \cap Y \notin \Gamma$, 
\begin{equation*}h(X) + h(Y) \geq h(X \cap Y) + h(X \cup Y) + 1. \end{equation*}
\end{itemize}

In order to prove the upper bound result, we first need a couple of lemmas.
For the sake of readability we abbreviate $h(\{x_1,\dots,x_i\})$ by $h(x_1..x_i)$.

\begin{la}\label{la:down} The function $h$ satisfies
\begin{equation*}\label{eq:down}
h(kp_1..p_{n-1})  \geq  h(p_1) + (n-1).
\end{equation*}
\end{la}

\begin{proof}
For any $i$ in the range $2 \leq i \leq n$, we have
$\{p_1,\dots,p_n\} \in \Gamma$, $\{k,p_1,\dots,p_{i-1},p_{i+1},\dots,p_n\} \in \Gamma$, and 
 $\{p_1,\dots,p_{i-1},p_{i+1},\dots,p_n\} \notin \Gamma$ (interpreting $p_{n+1},\dots,p_n$ as an empty sequence). 
By $+$-submodularity this produces the inequality 
\begin{equation*} h(p_1..p_n) + h(kp_1..p_{i-1}p_{i+1}..p_n) \geq h(p_1..p_{i-1}p_{i+1}..p_n) + h(kp_1..p_n) + 1. \end{equation*}
Adding the submodular inequality
\begin{equation*} h(p_1..p_i) + h(p_1..p_{i-1}p_{i+1}..p_n) \geq h(p_1..p_n) + h(p_1..p_{i-1})\end{equation*}
gives us
\begin{equation}\label{summand}
h(kp_1..p_{i-1}p_{i+1}..p_n) + h(p_1..p_i) \geq h(kp_1..p_n) + h(p_1..p_{i-1})+1.
\end{equation}

Summing (\ref{summand}) over $2 \leq i \leq n$, we get 
\begin{multline*}\displaystyle\sum_{i=2}^{n}h(kp_1..p_{i-1}p_{i+1}..p_n) + \displaystyle\sum_{i=2}^{n}h(p_1..p_i)  \geq\\
(n-1)h(kp_1..p_n) + \displaystyle\sum_{i=1}^{n-1}h(p_1..p_i) + (n-1). \end{multline*}

We now add the monotonicities $h(kp_1..p_n) \geq h(kp_1..p_{i-1}p_{i+1}..p_n)$ for $2 \leq i \leq n-1$ (not $n$) 
and $h(kp_1..p_n) \geq h(p_1..p_n)$.  Cancelling terms gives us the desired result.
\end{proof}


\begin{la}\label{la:up}The function $h$ satisfies the inequality
\begin{equation*}\label{eq:up}
h(kp_1) + \displaystyle\sum_{i=2}^{n-1}h(p_i) \geq h(kp_1..p_{n-1}) + (n-2).
\end{equation*}
\end{la}

\begin{proof}
For any $i$ in the range $2 \leq i \leq n-1$, we have
$\{kp_1\} \in \Gamma$,\\
 $\{kp_i\} \in \Gamma$, and $\{k\} \notin \Gamma$.  By +-submodularity this produces the inequality
\begin{equation*}h(kp_1) + h(kp_i) \geq h(kp_1p_i) + h(k) +1. \end{equation*}
Adding this to the submodular inequalities 
\begin{equation*} h(k) + h(p_i) \geq h(kp_i) \end{equation*}
and 
\begin{equation*} h(kp_1..p_{i-1}) + h(kp_1p_i) \geq h(kp_1..p_i) + h(kp_1) \end{equation*}
gives us 
\begin{equation}\label{summand2} h(kp_1..p_{i-1}) + h(p_i) \geq h(kp_1..p_i) + 1.
\end{equation}

Summing (\ref{summand2}) over $2 \leq i \leq n-1$, we get
\begin{equation*}
\displaystyle\sum_{i=1}^{n-2}h(kp_1..p_{i}) + \displaystyle\sum_{i=2}^{n-1}h(p_i) \geq \displaystyle\sum_{i=2}^{n-1}h(kp_1..p_i) + (n-2)
\end{equation*}
which simplifies to the desired result.
\end{proof}

\begin{thm}\label{upperbound}
For the king and $n$-pawn access structure $\Gamma_n$ 
\begin{equation*}\rho(\Gamma_n) \leq \frac{n-1}{2n-3}. \end{equation*}
\end{thm}

\begin{proof}
From Lemma \ref{la:down} we have 
\begin{eqnarray*}h(kp_1..p_{n-1})  &\geq&  h(p_1) + (n-1)\end{eqnarray*}
and from Lemma \ref{la:up} we have
\begin{eqnarray*}h(kp_1) + \displaystyle\sum_{i=2}^{n-1}h(p_i) &\geq& h(kp_1..p_{n-1}) + (n-2).\end{eqnarray*}
We add to these inequalities the submodularity
\begin{eqnarray*} h(p_1) +h(k) &\geq& h(kp_1)\end{eqnarray*}
to obtain
\begin{eqnarray*}  h(k) +\displaystyle\sum_{i=2}^{n-1}h(p_i) \geq  (2n-3). \end{eqnarray*}
Since
\begin{equation*}
h(k) +\displaystyle\sum_{i=2}^{n-1}h(p_i) \leq (n-1)\max_{p \in \{k,p_{2}..p_{n-1}\}} h(p)
\end{equation*}
at least one participant $p$ must satisfy $h(p) \geq \frac{2n-3}{n-1}$; equivalently,\\  
$\rho(p) \leq \frac{n-1}{2n-3}$.
Thus any secret sharing scheme for $\Gamma_n$ must have information rate at most $\frac{n-1}{2n-3}$,
and we conclude $\rho(\Gamma_n) \leq \frac{n-1}{2n-3}$.
\end{proof}

Next we prove the lower bound result.  By the definition of information rate for an access structure, it is sufficient to construct a secret sharing scheme $\Sigma$ for $\Gamma_n$ realizing
$\rho (\Sigma) = \frac{n-1}{2n-3}$. 
We will begin by exhibiting two schemes for $\Gamma_n$, $\Sigma_1$ and $\Sigma_2$, neither of which will attain the desired information rate.  We will then take $\Sigma$ to be a suitable weighted average of $\Sigma_1$ and $\Sigma_2$.
 
For the constructions of $\Sigma_1$ and $\Sigma_2$ we assume  the secret, $s$, is selected from the finite field $\mathbb{Z}_q$ where $q$ is a prime greater than $(2n-1)$.  Without loss of generality we may assume the secret is chosen with the uniform distribution \cite{BeiLiv}.

For the scheme $\Sigma_1$ we take a variant of an $(n, 2n-1)$ {\it threshold scheme} from \cite{Shamir}, which we describe here.  To share the secret $s$ we choose uniformly
at random a polynomial $f(x)$ over $\mathbb{Z}_q$ of degree at most $n-1$ with $f(0)=s$.   The king then receives as his $\Sigma_1$-share the values $f(1), f(2), \dots , f(n-1)$, and each $p_i$ for $1 \leq i \leq n$ receives the value $f(n-1+i)$.  That $\Sigma_1$ is indeed a secret sharing scheme for $\Gamma_n$ follows from the discussion of threshold schemes in \cite{Shamir}.  

\begin{la}\label{la:sigma1}
Each participant's $\Sigma_1$-share will occur with uniform distribution over the appropriate domain.
\end{la}

\begin{proof}
First, we observe that any $n$ equations of the form
$f(x_1)=y_1$ ,\dots, $f(x_n)=y_n$ for distinct values $x_1, \dots, x_n$ are satisfied by exactly one polynomial $f$ of degree at most $n-1$.  
For $m < n$ such equations $f(x_1)=y_1, \dots, f(x_m)=y_m$, there will be exactly $q^{n-m}$ polynomials that satisfy the set of equations. 
 To see this, fix $x_{m+1}, \dots , x_n$ distinct from each other and from the first $m$ values of $x$.  There are $q^{n-m}$ ways to choose values $y_{m+1},\dots,y_n \in \mathbb{Z}_q$, and each choice of values will yield a unique polynomial.  

We also observe that since the secret $s$ was chosen uniformly, and the polynomial $f$ of degree at most $n-1$ was chosen uniformly given the constraint $f(0)=s$, it follows that all polynomials $f$ over $\mathbb{Z}_q$ of degree at most $n-1$ are equally likely to be used in $\Sigma_1$.

From the above observations, we see that any possible $(n-1)$-tuple of values for the king will be realized by $q$ distinct polynomials.  Since each polynomial is equally likely, the king's $\Sigma_1$-share will be uniformly distributed over $\mathbb{Z}_q^{n-1}$.  Similarly, any pawn's share will be realized by $q^{n-1}$ distinct polynomials, each equally likely, and so a pawn's share will be uniformly distributed over $\mathbb{Z}_q$.
\end{proof}

The scheme $\Sigma_2$ is created using the decomposition method \cite{Stinson}, but the resulting scheme is simple enough to describe 
directly.  We use a $(2,2)$ threshold scheme and an $(n,n)$ threshold scheme, also discussed in \cite{Stinson}.  For the $(2,2)$ threshold scheme, we distribute shares to members of the minimal
qualified sets $\{k,p_i\}$ for $1 \leq i \leq n$, giving the king a random share $r\in \mathbb{Z}_q$, and giving each pawn the modular sum $r+s$.
For the $(n,n$) threshold scheme, we distribute shares to members of the remaining minimal qualified set $\{p_1, \dots , p_n\}$ by giving random values $r_i$, 
$1 \leq i \leq n-1$ to the first
$n-1$ pawns, and giving to $p_n$ the share
\begin{equation*}s + \displaystyle\sum_{i=1}^{n-1}r_i.\end{equation*}
The king's $\Sigma_2$-share will be just his share from the $(2,2)$ scheme.  A pawn's $\Sigma_2$-share will consist of his shares from both the $(2,2)$ and $(n,n)$ schemes.

Any qualified set in $\Gamma_n$ will be qualified in either the $(2,2)$ or the $(n,n)$ threshold scheme, and so will be qualified under $\Sigma_2$.  Any set $X \subset P_n$ not in $\Gamma_n$ will not be qualified in either of the threshold schemes being used for $\Sigma_2$.  If all random numbers are chosen independently and uniformly, $X$ will be unqualified under $\Sigma_2$.  Furthermore, the $\Sigma_2$ shares for the king and each pawn will be uniformly distributed over $\mathbb{Z}_q$ and $(\mathbb{Z}_q)^2$, respectively.

We are now ready to construct the scheme $\Sigma$, which will allow us to share a secret that consists of $n-1$ secrets from $\mathbb{Z}_q$.  To do this we share one secret over $\Sigma_1$, and one secret over each of $n-2$ copies of $\Sigma_2$.  We use uniformly and independently generated 
random numbers for each instantiation of a secret sharing scheme.

\begin{thm}\label{realize} 
\begin{equation*}\rho(\Sigma) = \frac{n-1}{2n-3}.\end{equation*}
\end{thm}

\begin{proof}
The king's $\Sigma$-share will consist of $(n-1) + (n-2) = 2n-3$ shares from $\mathbb{Z}_q$.  By Lemma \ref{la:sigma1} and the discussion regarding $\Sigma_2$ any $\Sigma$-share for the king will be equally likely, and so
\begin{equation*} \rho(k) = \frac{H(S)}{H(k)} = \frac{\log(q^{n-1})}{\log(q^{2n-3})} = \frac{n-1}{2n-3}. \end{equation*}
Similarly, each pawn's $\Sigma$-share will consist of $1 + 2(n-2)= 2n-3$ shares from $\mathbb{Z}_q$, all possible $\Sigma$-shares having equal likelihood, 
and so we also have 
\begin{equation*}\rho(p) = \frac{n-1}{2n-3}.\end{equation*}
Thus
\begin{eqnarray*}
\rho(\Sigma) &=& \min \{\rho(k), \rho(p)\} = \frac{n-1}{2n-3}.
\end{eqnarray*}
\end{proof}

\begin{thm}\label{lowerbound} For the king and n-pawn access structure $\Gamma_n$
\begin{equation*}\rho(\Gamma_n) \geq \frac{n-1}{2n-3}.\end{equation*}
\end{thm}

\begin{proof}
This follows immediately from Theorem \ref{realize} and the definition of information rate for an access structure.
\end{proof}

\begin{proof}[Proof of Theorem 1]
Combining the bounds found in Theorems \ref{upperbound} and \ref{lowerbound}, we conclude that
\begin{equation*}
\rho(\Gamma_n) = \frac{n-1}{2n-3}.
\end{equation*}
\end{proof}

\section{Acknowledgments}
The author would like to thank Andreas Blass for many helpful discussions, and Carles Padr\'{o} for providing an updated version of his paper \cite{MF} with Jaume Mart\'{i}-Farr\'{e}.

\end{document}